\documentclass[a4paper,twocolumn,shownopacs,,amsmath,amssymb,prl,groupedaddress]{revtex4-1}

\usepackage[draft]{hyperref}
\usepackage{amsmath}
\usepackage{graphicx}
\usepackage{dcolumn}
\usepackage{amssymb}
\usepackage{color}
\usepackage{anyfontsize}

\newcommand{\ba}{\begin{array}}
\newcommand{\ea}{\end{array}}
\newcommand{\be}{\begin{equation}}
\newcommand{\ee}{\end{equation}}
\newcommand{\bea}{\begin{eqnarray}}
\newcommand{\eea}{\end{eqnarray}}

\begin{document}

\title{Bragg-Berry mirrors: reflective broadband q-plates}

\author{Mushegh Rafayelyan and Etienne Brasselet*}

\address{$^1$Univ. Bordeaux, LOMA, UMR 5798, F-33400 Talence, France\\
$^2$CNRS, LOMA, UMR 5798, F-33400 Talence, France}

\begin{abstract}
We report on the experimental realization of flat mirrors enabling the broadband generation of optical vortices upon reflection. The effect is based on the geometric Berry phase associated with the circular Bragg reflection phenomenon from chiral uniaxial media. We show the reflective optical vortex generation from both diffractive and nondiffractive paraxial light beams using spatially patterned chiral liquid crystal films. The intrinsic spectrally broadband character of spin-orbit generation of optical phase singularities is demonstrated over the full visible domain. Our results do not rely on any birefringent retardation requirement and consequently foster the development of a novel generation of robust optical elements for spin-orbit photonic technologies.
\end{abstract}


\maketitle

Tailoring intensity, phase and polarization of electromagnetic fields is a cornerstone of today’s information and communication technologies. In particular, ever-increasing demand for higher information-rate has led photonic technologies to play a paramount role, especially through the use of the orbital angular momentum that light may carry and which can be associated with an infinite-dimensional basis. The most common illustration corresponds to the paraxial description of electromagnetic fields by means of so-called Laguerre-Gauss modes. Indeed, each of these modes is characterized by an optical phase singularity with integer topological charge $\ell$ that is described by a complex field amplitude proportional to $\exp(i\ell\phi)$ ($\phi$ is the usual azimuthal angle in the transverse plane) and associated with orbital angular momentum $\ell\hbar$ per photon \cite{allen_pra_1992} ($\hbar$ is the reduced Planck constant). Therefore, the production of singular phase mask having an amplitude transmittance proportional to $\exp(i\ell\phi)$ is a basic issue for which several technologies have been developed \cite{yao_aop_2011}, among which powerful ones based on helicity-dependent geometric phases \cite{vinitskii_ufn_1990, bhandari_physrep_1997}. In particular, the use of azimuthally varying anisotropic structures for helicity-controlled generation of optical vortex beams emerged in the early 2000s \cite{biener_ol_2002}. Namely, a circularly polarized paraxial light beam with helicity $\sigma=\pm1$ passing through a half-wave plate having optical axis orientation angle of the form $\psi = q\phi$ with $q$ half-integer (in short, a ``$q$-plate'') experiences a change of its orbital state $\ell$ by an amount $\Delta\ell = 2\sigma q$. Retrospectively, its significance started to be fully appreciated after nematic liquid crystals $q$-plates working in the visible domain have been realized in 2006 \cite{marrucci_prl_2006} and that the coupling between the spin and orbital angular momentum of light has been pointed out. Indeed, since then, $q$-plates eased and made possible numerous fundamental and applied realizations in optics, both in the classical and quantum regimes, as reviewed in \cite{marrucci_jo_2011}.

Importantly, the latter $q$-plates are transmissive spin-orbit optical elements that require $\pi$ birefringent phase retardation condition. In this context, we recently unveiled geometric Berry phase associated with the circular Bragg reflection phenomenon in chiral anisotropic optical media \cite{rafayelyan_prl_2016} that brings a novel paradigm to achieve wavelength-independent pure spin-orbit topological shaping of light from two- and three-dimensional ``Bragg-Berry'' mirrors (i) in the reflection mode, (ii) without need for any birefringent retardation requirement. Existence of a geometric phase at reflection from cholesteric liquid crystals planar slabs is also discussed in another recent independent report \cite{barboza_arxiv_2016}. From the application point of view, we also note a closely related recent work introducing the control of reflected wavefronts from planar optics with patterned cholesteric liquid crystals \cite{kobashi_np_2016} with the realization of reflective lenses and deflectors from Bragg-Berry mirrors.

Here we propose, realize and experimentally demonstrate the broadband reflective optical vortex generation from flat Bragg-Berry mirrors for both diffractive and nondiffractive paraxial light beams. This is done by using either Gaussian or Bessel incident beams whose reflection off a chiral Bragg mirror endowed with suitable surface orientational boundary conditions lead to the production of Laguerre-Gauss and higher-order Bessel like beams, respectively. From the spectral point of view, we report on optical vortex generation over the full visible range, at least from 450 to 650~nm wavelengths. By doing so, our results substantially extend the reach of previous works \cite{rafayelyan_prl_2016, barboza_arxiv_2016, kobashi_np_2016} and set the basis for the development of a novel generation of spin-orbit optical elements whose demonstrated robustness against polychromaticity and fabrication constraints offer a valuable alternative to existing $q$-plates limitations. Still, broadband transmissive liquid crystal $q$-plates of arbitrary order can be realized using chiral multilayer structures \cite{li_spie_2012}. Solid-state strategies realizing effective $q$-plate endowed with broadband capabilities have also been reported \cite{bouchard_njp_2014, radwell_nc_2016}. However, the latter approaches are limited to $q=1$ and the poor spatial resolution of the effective optical axis patterning (several millimeters) limits their application potential.

In practice, cholesteric liquid crystals are prime choice materials to realize chiral Bragg mirrors. In fact they combine the strong optical anisotropy, $\Delta n$, of nematic liquid crystals with a helical supramolecular ordering of the liquid crystal molecules characterized by a pitch, $p$, that is the distance over which the average molecular axis rotates by $2\pi$. As such, thick enough cholesteric films (a few pitches are enough) may selectively reflect one of two circular polarization states via the circular Bragg reflection phenomenon \cite{faryad_aop_2014}. Considering light propagation along the helical axis, the circular Bragg reflection corresponds to the existence of a photonic bandgap centered on the wavelength $\lambda = np$ ($n$ is the average refractive index of the cholesteric) and characterized by a spectral width $\Delta\lambda = p\Delta n$ \cite{oswald_book_nematic}. The Bragg-reflected photons are those having the helicity $\sigma = -\chi$, where $\chi=\pm1$ is the handedness of the supramolecular cholesteric helix. As shown in Ref.~\cite{rafayelyan_prl_2016}, it is the preservation of the photon helicity at circular Bragg reflection, associated with the molecular orientation at the surface of the mirror, that leads to the production of a geometric Berry phase for the reflected light. More precisely, previous works established that a cholesteric Bragg mirror characterized by a surface molecular orientation angle $\psi$ generates a phase factor $2\sigma\psi$ for the Bragg-reflected field \cite{rafayelyan_prl_2016, barboza_arxiv_2016, kobashi_np_2016}. We thus expect that Bragg-Berry mirrors having a patterned surface molecular orientation of the form $\psi = q\phi$, $q$ half-integer, produce a change of the incident orbital state $\ell$ by an amount $\Delta\ell = -2\chi q$.

Flat Bragg-Berry mirrors are prepared from a $10~\mu$m-thick right-handed cholesteric Bragg mirror made of liquid crystal material MDA-02-3211 (from Merck), that is characterized by pitch $p=347$~nm, birefringence $\Delta n=0.195$ and average refractive index $n=1.604$ at 589.3 nm wavelength and temperature $20^\circ$C. The cholesteric slab is sandwiched between two glass substrates coated by an azobenzene-based surface-alignment layer with submicron thickness (from Beam Co.), as depicted in Fig.~1(a). Two alignment layers are optically treated to ensure azimuthally varying orientation of the liquid crystal molecules at both ends of the liquid crystal layer of the form $\psi = q\phi$ with $q$ half-integer. In this work, without loss of generality, we restrict our demonstration of principle to the case $q=1$. This is illustrated in Fig.~1(c) that shows the spatial distribution of the orientational boundary conditions obtained by polarization microscopy (CRi Abrio Micro imaging system) of a coated glass substrate alone. The residual birefringent phase retardation of the the alignment layer is of the order of $0.01\pi$ and its contribution to the reflected field is weak enough to be safely discarded.

\begin{figure}[t!]
\centering\includegraphics[width=1\columnwidth]{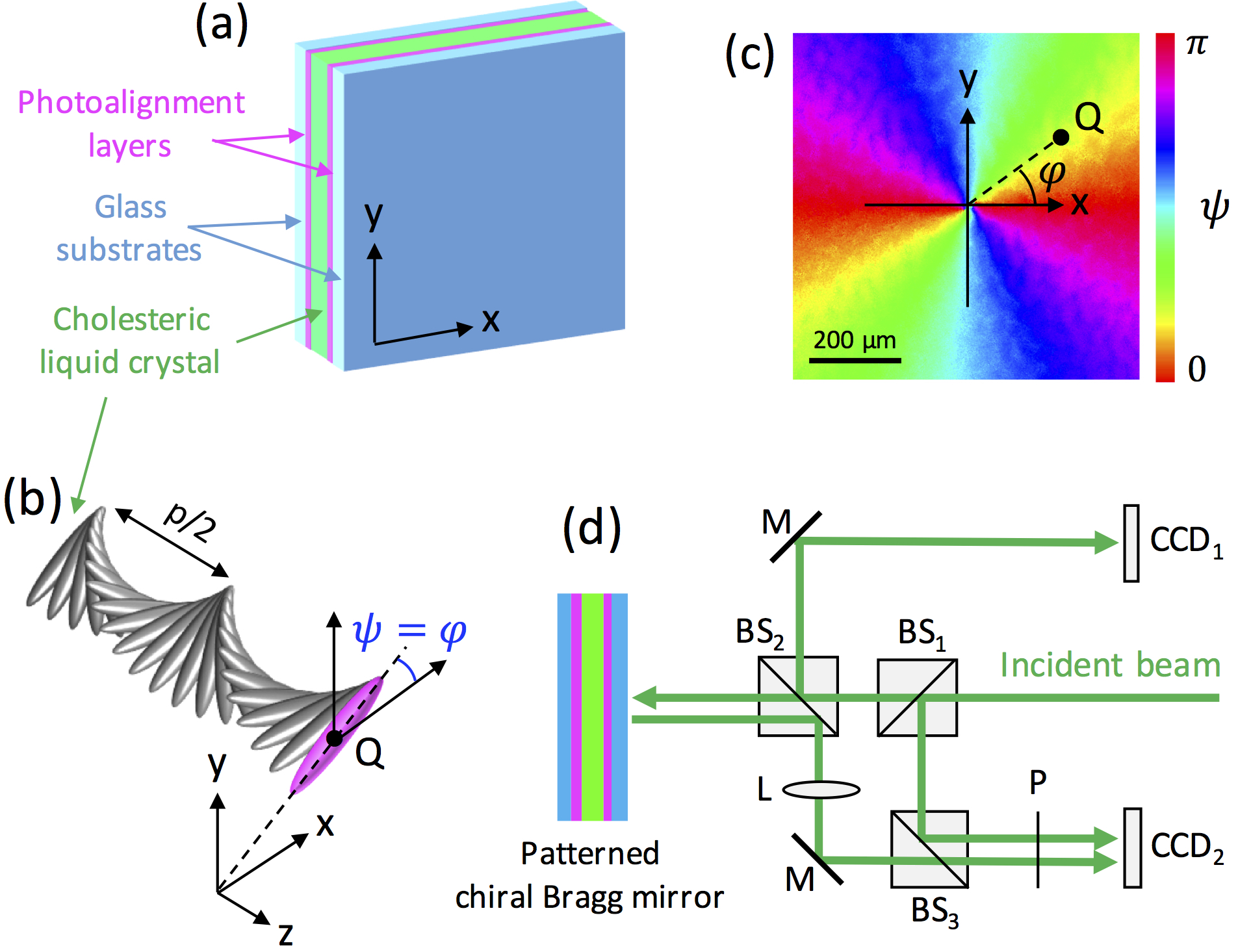}
\caption{
(a) Sketch of the Bragg-Berry flat mirror that consists of a space-variant cholesteric liquid crystal flat mirror sandwiched between two planar glass substrates provided with photoalignment layers that ensure radial ordering of the liquid crystal molecules at both ends of the liquid crystal slab. (b) Illustration of the cholesteric liquid crystal supramolecular right-handed ($\chi=1$) helix with pitch $p$ that is oriented along the $z$ axis. The radial surface anchoring condition is visualized by the ``molecule'' at $z=0$ in magenta color. (c) Experimental map of the space-variant orientation $\psi$ of the alignment layers in the $(x,y)$ plane. (d) Experimental setup for optical vortex generation from an incident beam impinging at normal incidence on the Bragg-Berry mirror. M, mirror; BS$_{1,2,3}$, beamsplitters; L, lens; P, polarizer; CCD$_{1,2}$, imaging devices.
}
\end{figure}

Experimental demonstration of optical vortex generation is done at 532~nm wavelength (that falls in the circular photonic bandgap of our material) by using a continuous laser beam, following the setup shown in Fig.~1(d). First, the Bragg-Berry mirror is illuminated at normal incidence by a circularly polarized Gaussian beam with helicity $\sigma=-\chi$ with waist radius in the plane of the mirror $w \sim 0.25~$mm, see Fig.~2(a). The far-field reflected light is collected by camera CCD$_2$ by placing the lens L in $f$-$f$ configuration with respect the patterned chiral Bragg mirror and the camera, see Fig.~2(b). A doughnut intensity profile characteristic of the presence of the expected on-axis optical phase singularity of topological charge $-2\chi$ is observed. Theoretically this is described by Fourier transform, which expresses in the polar coordinate $(k,\theta)$ as $E(k,\theta) \propto \int_0^{2\pi}\int_0^\infty \Phi(\phi)G(r) \exp[-i k r \cos(\theta-\phi)] r dr d\phi$ where $G(r) =\exp(-r^2/w^2)$ is the Gaussian beam incident amplitude on the mirror and $\Phi(\phi)=\exp(- 2i\chi \phi)$ is the reflective phase mask of the Bragg-Berry mirror. Integration over the azimuthal coordinate gives $E(k,\theta) \propto \exp(- 2i\chi \theta) \int_0^\infty G(r) J_{-2\chi}(k r) r dr$ where $J_n$ is the $n$th-order Bessel function of the first kind, which unveils the optical phase singularity. The corresponding intensity profile is axisymmetric and is given by $I(k) \propto [1-\exp(-k^2w^2/4)(1+k^2w^2/4)]/k^2$. The result of simulations is shown in Fig.~2(c) that displays both the intensity (luminance) and phase (colormap) profiles, which gives fair agreement.

\begin{figure}[t!]
\centering\includegraphics[width=1\columnwidth]{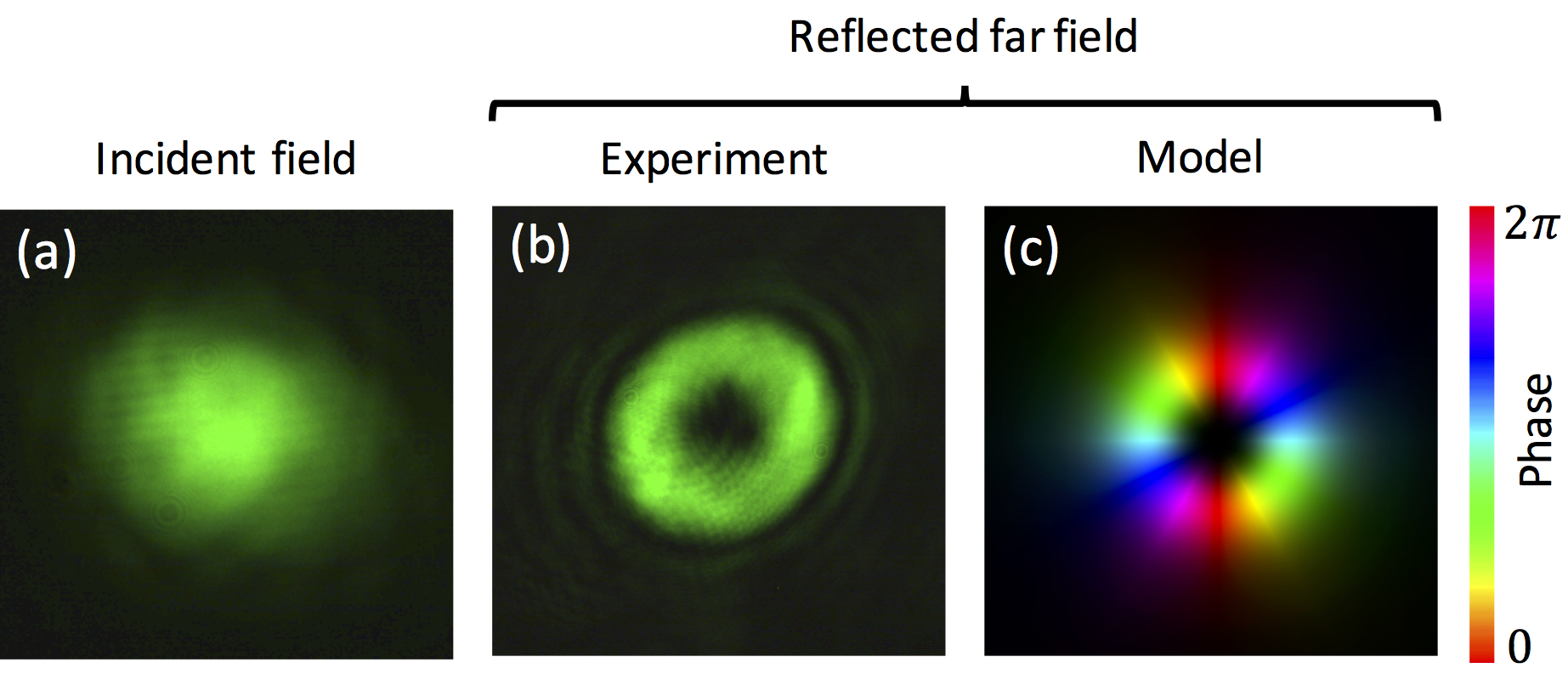}
\caption{
(a) Experimental incident Gaussian intensity profile on the sample collected by camera CCD$_1$. (b) Far field intensity profile collected by camera CCD$_2$ owing to the lens L placed in $f$-$f$ configuration. (c) Calculated intensity and phase that correspond to panel (b). The luminance refers to the intensity and the colormap refers to the phase from $0$ to $2\pi$.
}
\end{figure}

\begin{figure}[b!]
\centering\includegraphics[width=1\columnwidth]{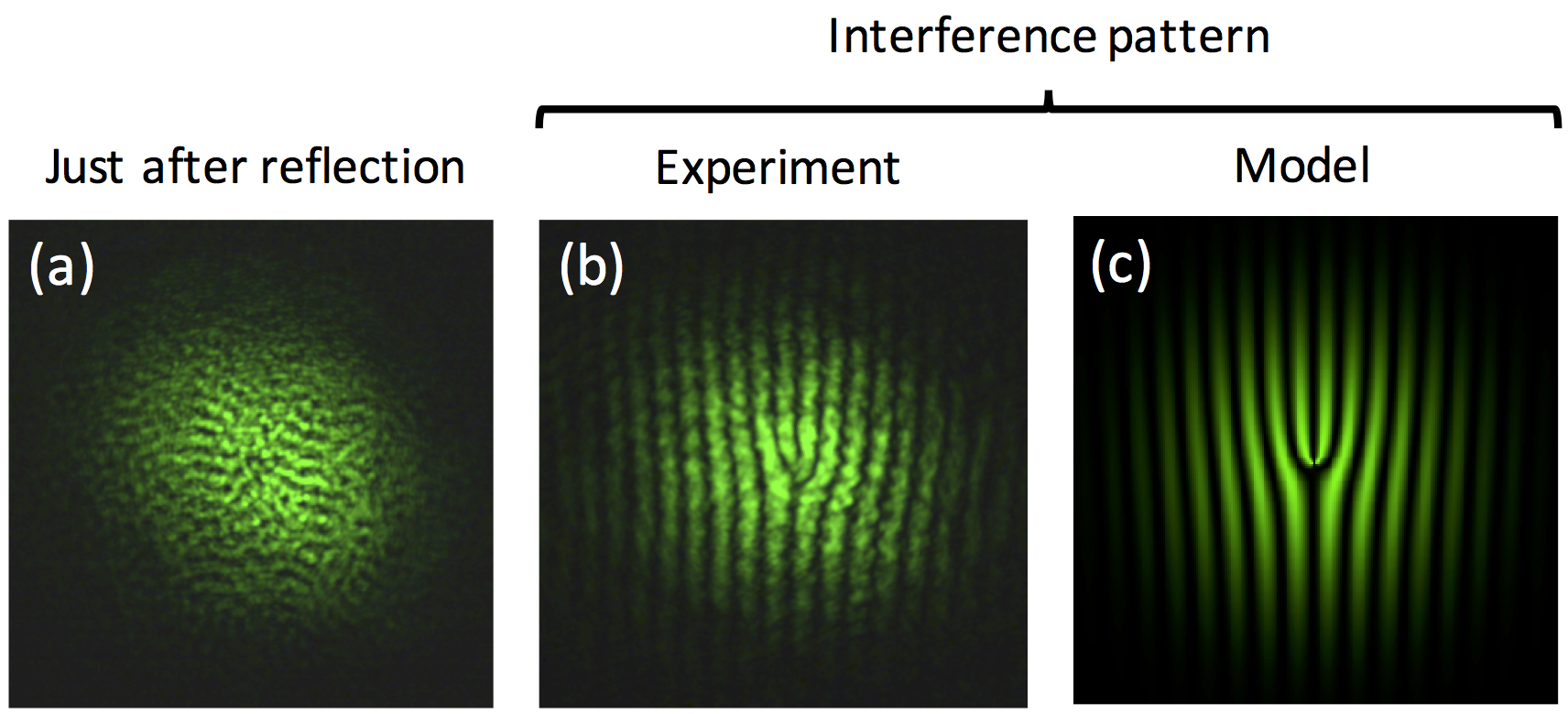}
\caption{
(a) Experimental intensity profile just after a normally incident Gaussian beam is reflected off the Bragg-Berry mirror, that is collected by camera CCD$_2$ placing the lens L in $2f$-$2f$ configuration. The speckle-like pattern is reminiscent of residual imperfections of the alignment layer. (b) Interference pattern between the field that corresponds to panel (a) and a reference Gaussian beam. A fork-like interference pattern of order two is observed. (c) Simulation of the observed singular interference pattern shown in panel (b).
}
\end{figure}

Further support of singular beam shaping dictated by a change $\Delta\ell = -2\chi q$ of the incident orbital state is made by analyzing the phase profile of the field just after it reflects off the sample. This is shown in Fig.~3(a) that displays the intensity pattern collected by camera CCD$_2$ by placing the lens L in $2f$-$2f$ configuration with respect to the sample and the camera. Obviously, the observed intensity pattern corresponds to a Gaussian enveloppe since diffraction has not yet taken place. However, an optical phase singularity is already imprinted into the field, which is revealed by interfering with the field with a reference Gaussian beam. The resulting interference pattern is shown in Fig.~3(b) that exhibits a fork-like intensity pattern with two teeth, whose intensity modulation contrast is optimized by placing a polarizer before CCD$_2$ (see Fig.~1(d)). This demonstrates experimentally that the mirror behaves as effective reflective singular phase mask with topological charge 2 as indicated above while describing the reflected far field. The latter ``near field'' observations are confronted with simulations in Fig.~3(c) that displays the intensity pattern $|\Phi G + G_{\rm ref}|^2$ where $G_{\rm ref}$ refers to the reference Gaussian beam whose propagation direction is slightly tilted with respect to that of the probed field.

\begin{figure}[t!]
\centering\includegraphics[width=1\columnwidth]{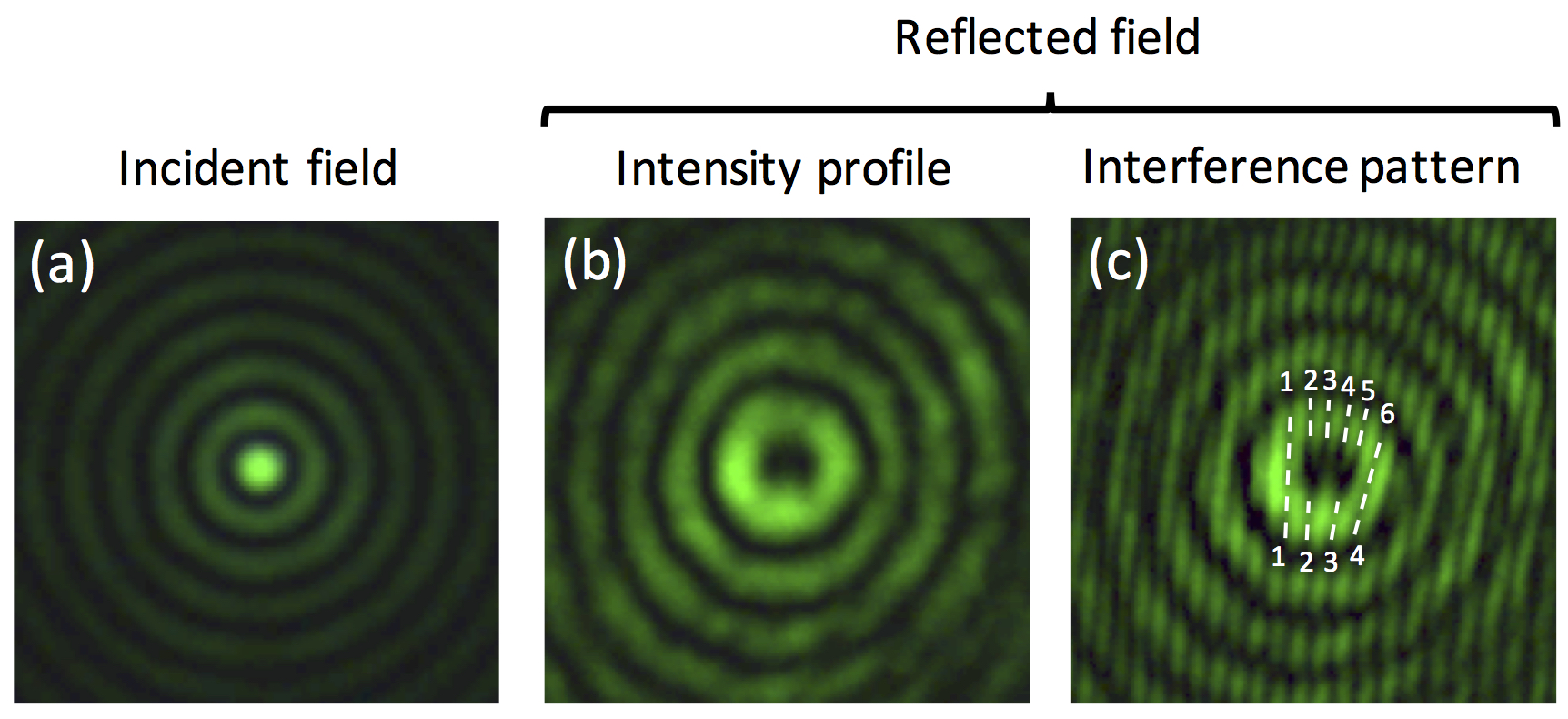}
\caption{
(a) Experimental incident fundamental Bessel intensity profile on the sample collected by camera CCD$_1$. (b) Reflected field intensity profile collected by camera CCD$_2$. (c) Interference pattern between the field that corresponds to panel (b) and a reference Gaussian beam. A fork-like interference pattern of order two is observed. Dashed segments and fringes numbering are given as a guide.
}
\end{figure}

\begin{figure}[b!]
\centering\includegraphics[width=1\columnwidth]{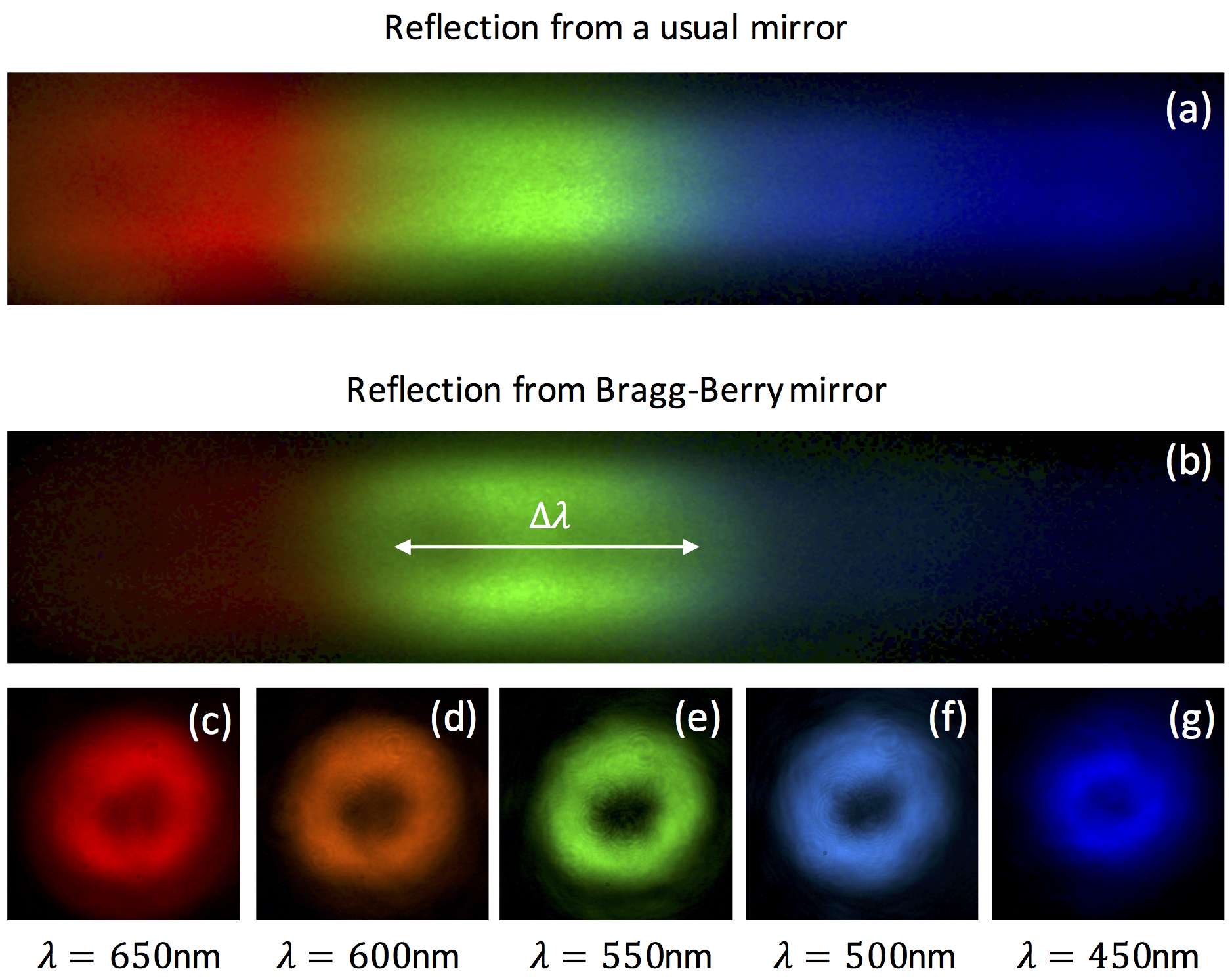}
\caption{
(a) White-light laser source spectrum of the reflected beam off a usual broadband dielectric mirror using a dispersion prism. (b) Same as panel (a) in the case a Bragg-Berry mirror with $q=1$. (c) to (g) Selection of far field vortex beam intensity profiles for a set of wavelengths.
}
\end{figure}

\begin{figure}[t!]
\centering\includegraphics[width=1\columnwidth]{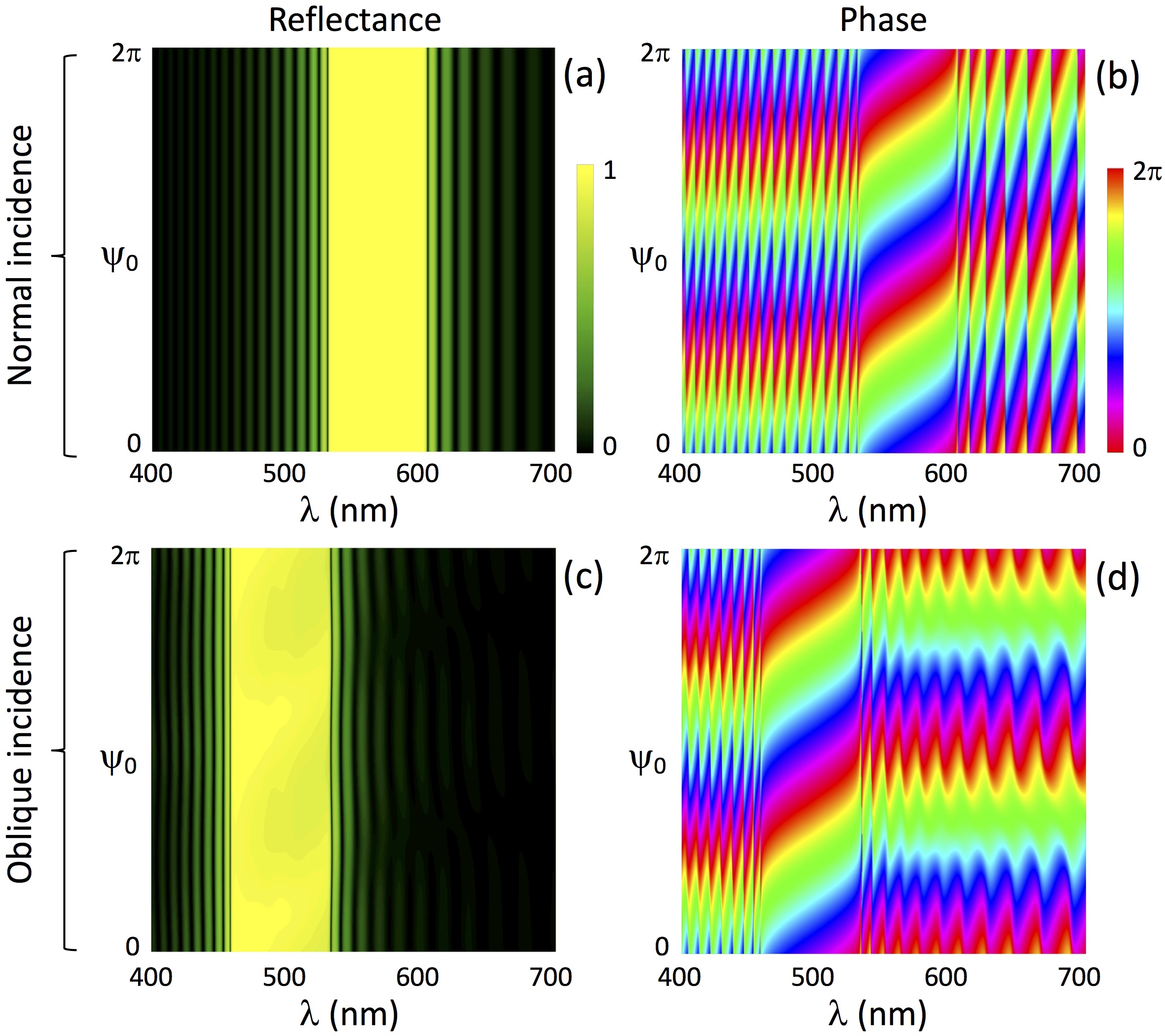}
\caption{
(a,b) Reflectance and phase spectra of the helicity-preserved liqght field component of the reflected field for Bragg-polarized incident field as a function of $\psi_0$. (c,d) Same as in (a) and (b) for $30^\circ$ incident angle in the glass substrate. Calculations are made using the refractive indices parallel and perpendicular to the director of our liquid crystal mixture, namely $n_\parallel = 1.7013$ and $n_\perp = 1.5064$ at 589.3 nm wavelength and temperature $20^\circ$C, and glass substrate refractive index $n_{\rm glass}=1.52$.
}
\end{figure}

Then we formally extend above topological shaping of light to diffraction-free optical fields. This is done by preparing the incident field as a circularly polarized fundamental Bessel beam with helicity $\sigma=-\chi$, placing a glass axicon in the course of above incident Gaussian beam \cite{mcgloin_cp_2005}, between beamsplitters BS$_1$ and BS$_2$ (see Fig.~1(d)). The incident intensity pattern collected by camera CCD$_1$ is shown in Fig.~4(a), which corresponds to an axicon with full-apex cone angle of $178^\circ$. The reflected field after a few centimeters of propagation is imaged on CCD$_2$ and shown in Fig.~4(b). The corresponding intensity pattern is representative of the expected higher-order Bessel beam as expected from the reflective phase mask $\rho(\phi)$, whose order two is unveiled by interferometry as shown in Fig.~4(c) where a tilted Gaussian reference arm is used. Indeed, by counting the fringes around the central part of the pattern, one can retrieve the characteristic $4\pi$ exploration of the phase along the circle.

Finally, we explore the polychromatic behavior of the Bragg-Berry mirror in the visible domain by replacing the monochromatic laser beam by a supercontinuum whose spectrum is presented in Fig.~5(a). It corresponds to the image of the reflected beam from a usual broadband mirror using a dispersion prism. Replacing the latter mirror by the Bragg-Berry mirror, one gets the Bragg reflection spectrum shown in Fig.~5(b) that exhibits the characteristic high reflectance spectral bandwidth $\Delta \lambda$ of a cholesteric mirror. However, the singular nature of Bragg-reflected field goes beyond the spectral region of the circular photonic bandgap. This is illustrated in the bottom part of Fig.~5 that shows the intensity patterns of the Bragg-reflected polychromatic vortex beam with topological charge 2 for a set of five wavelength from 450 to 650~nm by step of 50~nm, following the monochromatic approach used to obtain Fig.~2(b) by using a set spectral filters with 10~nm full-width half-maximum placed before the camera CCD$_2$. As one can see from Figs.~5(c) to 5(g), the quality of the generated optical beam is decreased outside the photonic bandgap, where the reflected signal is low, and several factors can alter topological shaping purity, such as the imperfect refractive index matching conditions between glass substrate and the average refractive index of the anisotropic medium \cite{faryad_aop_2014} and the presence of photoalignment layers. In practice, this can be easily circumvented by using either external fields (thermal, electrical, magnetic) effects or oblique incidence that are well-known to drastically shift the bandgap \cite{oswald_book_nematic}.

Although the reflected intensity is obviously wavelength-dependent outside the circular photonic bandgap, the purity of the generated vortex is wavelength-independent. This is illustrated in Figs.~6(a) and 6(b) where the reflectance and phase spectra of the helicity-preserved reflected field for incident Bragg-polarized light is calculated from Berreman approach \cite{berreman_josa_1972} as a function of the orientation, say $\psi_0$, of the liquid crystal director at $z=0$ (see Fig.~1). Therefore, our approach does not require the use of post polarization filtering as is the case of usual transmissive $q$-plate. In our case, low reflectance outside bandgap implies to use larger acquisition time, hence lower signal-to-noise ratio, as one can see in Figs.~5(c) and 5(g). Moreover, the robustness of the pure broadband optical vortex generation versus the angle of incidence is illustrated in Figs.~6(c) and 6(d) for $30^\circ$ external incidence angle in glass substrate. Note that the two main effects of oblique incidence are (i) to shift the bandgap to the blue spectral region and (ii) to break axisymmetry even more than the incidence angle increases, as one can see from the modulated reflectance versus $\psi_0$ in Fig.~6(c). This emphasizes the large acceptance angle of Bragg-Berry mirrors.

Summarizing, we have demonstrated that flat spin-orbit reflective optical elements enable complex phase engineering of light fields, whatever their diffractive or non-diffractive nature. Noticeably, existing strategies to go beyond present limitations associated with helicity dependent high reflectance regime over a spectral window of width of a few tens of nanometers, which had been originally developed for non-patterned cholesteric mirrors as reviewed in \cite{mitov_advmat_2012}, could formally be extended to the case of patterned surface anchoring conditions. In general, the use of anisotropic chiral metamaterials may offer exotic photonic bandgaps  \cite{gevorgyan_jo_2013}. Moreover, the sensitivity of liquid crystals to external fields should bring tunable or self-induced operating conditions. On the other hand, state-of-the-art point-by-point photoalignment techniques allow considering arbitrary singular or non-singular patterns \cite{kim_optica_2015, ji_scirep_2016}. Finally, recalling that Bragg-Berry mirrors have no restrictions in terms of cell thickness and that curvature may lead to additional topological shaping features \cite{rafayelyan_prl_2016}, this work should foster the development of spin-orbit photonic technologies.

\vspace{2mm}
\noindent
*etienne.brasselet@u-bordeaux.fr

\vspace{2mm}
\noindent
M.R. acknowledges the financial support from Erasmus Mundus Action 2 BMU-MID under Grant MID20121729.


\end{document}